\documentclass[]{interact}

\usepackage{epstopdf} 
\usepackage[caption=false]{subfig} 

\usepackage[numbers,sort&compress]{natbib} 
\bibpunct[, ]{[}{]}{,}{n}{,}{,} 
\renewcommand\bibfont{\fontsize{10}{12}\selectfont} 

\theoremstyle{plain} 

\theoremstyle{definition}

\theoremstyle{remark}

\usepackage{amsmath}
\usepackage{amssymb}
\usepackage{dsfont}
\usepackage[dvipsnames]{xcolor}  

\usepackage{mathtools} 
\newtagform{eqlabcolour}[\color{purple}]{\color{purple}(}{)}

\begin{document}

\articletype{APPLICATION NOTE} 

\title{Compositional data analysis for modelling and forecasting mortality using the $\alpha$-transformation}

\author{
\name{Han Ying Lim\textsuperscript{a}, Dharini Pathmanathan\textsuperscript{a,b,c}\thanks{CONTACT Dharini Pathmanathan. Email: dharini@um.edu.my} and Sophie Dabo-Niang\textsuperscript{d,e}}
\affil{\textsuperscript{a}Institute of Mathematical Sciences, Faculty of Science, Universiti Malaya, 50603 Kuala Lumpur, Malaysia; \textsuperscript{b}Universiti Malaya Centre for Data Analytics, Universiti Malaya, 50603 Kuala Lumpur, Malaysia; \textsuperscript{c}Center of Research for Statistical Modelling and Methodology, Faculty of Science, Universiti Malaya, 50603 Kuala Lumpur, Malaysia; \textsuperscript{d}CNRS, UMR 8524-Laboratoire Paul Painlev\'{e}, INRIA-MODAL, Universit\'{e} Lille, F-59000 Lille, France; \textsuperscript{e}CNRS-Universit\'{e} de Montr\'{e}al, CRM-CNRS, Montr\'{e}al, Canada.}
}

\maketitle


\begin{abstract}
Mortality forecasting plays a central role in both demographic and actuarial studies. Classical approaches, such as the Lee-Carter (LC) model, typically rely on mortality rates as the primary measure. In recent years, compositional data analysis (CoDA), which respects summability and non-negativity constraints, has gained increasing attention for mortality forecasting. While the centred log-ratio (CLR) transformation is commonly used to map compositional data to real space, the $\alpha$\nobreakdash-transformation, a generalisation of log-ratio transformations, offers greater flexibility and adaptability. This study introduces the $\alpha$\nobreakdash-transformation as an alternative to the CLR transformation within a CoDA framework under a discrete-age setting for all-cause mortality forecasting, which has not been previously investigated. To enable a fair comparison of transformation choices, zero values in the data are imputed, although the $\alpha$\nobreakdash-transformation can inherently accommodate them. Using age-specific life table death counts for males and females in 31 selected European countries/regions from 1983 to 2018, the proposed method demonstrates comparable performance to the CLR transformation in most cases, with improved forecast accuracy in some instances. These findings highlight the potential of the $\alpha$\nobreakdash-transformation to enhance mortality forecasting within the CoDA framework.
\end{abstract}



\section{Introduction}

Mortality forecasting plays a central role in both demographic and actuarial studies, informing strategic planning for national healthcare systems across the entire life course and supporting assessments relevant to pension design and longevity-related financial products \cite{HLSAlpha2024, Bas2020}. While classical mortality modelling often relies on central mortality rates \cite{Hyn2007, Lee1992, Lee2001, Li2005}, there is increasing recognition of the importance of forecasting age-at-death distributions over time \cite{Bgr2017, Gia2024, Kj2019, Oep2008, HLS2020, HLSWeighted2024, HLSAlpha2024}, as they provide a more comprehensive representation of mortality experience within a population \cite{Pas2019} and contain measures of central tendency for longevity \cite{Can2010}. By characterising how deaths are gradually redistributed from younger to older ages, life table death counts $d_{t,x}$ capture age-structured mortality dynamics and describe patterns related to lifespan variability at the population level \cite{Abu2020, Pas2019}.

The development of mortality forecasting methods dates back to the twentieth century \cite{Pol1988}. In general, these methods can be categorised into expert judgement, extrapolative and epidemiological approaches \cite{Booth2008}. Among these, the Lee-Carter (LC) model, which models log mortality rates, has gained widespread popularity since its introduction \cite{Lee1992} and has been extended into various variants to improve forecast accuracy \cite{Bas2023}. The classical LC model employs a statistical time series approach to project a single time-varying parameter for mortality forecasting, making it a primarily extrapolative method with minimal subjective judgment \cite{Bas2023}. The time index $\kappa_t$ and age pattern $\beta_x$ are estimated using singular value decomposition (SVD) applied to a centred matrix of log mortality rates, enabling the projection of life table death densities \cite{Hyn2007, Lee1992, Li2005}. This approach leverages the approximately log-linear decline in age-specific mortality over time and allows the use of multivariate statistical techniques designed for unbounded variables \cite{Oep2008}.

Compositional data analysis (CoDA) is an analytical framework for handling compositional data, which are positive vectors carrying relative information that represent parts of a whole with a fixed sum, such as proportions \cite{Ait1986}. Such data frequently arise in fields like geochemistry and atmospheric science. Standard statistical analyses require compositional data to be transformed from the Aitchison simplex to real space \cite{Ait1986}. The centred log-ratio (CLR) transformation is widely used due to its interpretability and ability to preserve distances \cite{Ait1986}. Since $d_{t,x}$ values are non-negative, range between 0 and the life table radix, and sum naturally to the radix each year \cite{Oep2008}, they can be treated as compositional data. Forecasting $d_{t,x}$ using a log-linear approach often produces predicted values that vary independently across ages and fail to preserve the life table radix constraint \cite{Bgr2019}. This limitation can be addressed within the CoDA framework, which leverages the constant-sum property to induce a natural covariance structure among components \cite{Bgr2019}.

Oeppen’s pioneering work \cite{Oep2008} introduced a CoDA-based framework for mortality forecasting, analogous to the LC model, which focuses on the redistribution of $d_{t,x}$. Oeppen \cite{Oep2008} showed that multiple-decrement compositional forecasts by age and cause are not necessarily more pessimistic than single-decrement forecasts by age alone, contradicting earlier findings based on mortality rates \cite{Wil1995}. Bergeron-Boucher et al. \cite{Bgr2017} extended the CoDA model to enable regional coherent mortality forecasting, similar to the Li-Lee model \cite{Li2005}. Their findings highlight that both coherent and non-coherent CoDA models produce less biased forecasts with higher accuracy for many countries compared to LC-based counterparts. This improvement is partly attributed to using $d_{t,x}$ as the mortality measure and applying the CLR transformation, which accounts for the changing rate of mortality improvement over time \cite{Bgr2017}. Furthermore, the summability constraint inherent in compositional data preserves coherence across populations, addressing a key limitation of the LC model \cite{Bgr2017}. Recognising the strength of the CoDA framework in capturing dependencies between causes of death, Kjærgaard et al. \cite{Kj2019} proposed two CoDA-based models to forecast cause-specific death distributions within a single population. Shang and Haberman further extended the CoDA framework to a functional setting \cite{HLS2020, HLSWeighted2024}, adapting the Hyndman-Ullah (HU) model \cite{Hyn2007}. Collectively, these studies demonstrate that CoDA-based approaches can improve the accuracy of mortality forecasting.

Most CoDA-based studies rely on log-ratio transformations of $d_{t,x}$, which have notable drawbacks, such as they lack flexibility and cannot handle zeros due to their logarithmic nature. These limitations can be addressed by the $\alpha$\nobreakdash-transformation introduced by Tsagris et al. \cite{Tsa2011}, which generalises log-ratio approaches and offers greater flexibility through the $\alpha$\nobreakdash-parameter \cite{Tsa2011}. Typically ranging between 0 and 1 \cite{Gia2024, HLSAlpha2024}, $\alpha$ balances log-ratio analysis (LRA, $\alpha = 0$) and Euclidean data analysis (EDA, $\alpha = 1$), with intermediate values sometimes outperforming both \cite{Tsa2016}. Importantly, strictly positive $\alpha$ values allow the transformation to accommodate zero-containing data \cite{Tsa2011, Tsa2016, Tsa2023}. Although not all of Aitchison’s \cite{Ait1986} theoretical properties such as scale invariance, perturbation invariance and subcompositional dominance, which primarily support LRA methods, are satisfied by the $\alpha$\nobreakdash-transformation, its practical applicability remains unaffected \cite{Tsa2011, Tsa2023}.

Numerous empirical studies have demonstrated that the $\alpha$\nobreakdash-transformation can improve performance in both regression \cite{Tsa2015} and classification tasks \cite{Tsa2016}. In forecasting, Shang and Haberman \cite{HLSAlpha2024} found that the $\alpha$\nobreakdash-transformation outperformed log-ratio approaches within the functional CoDA framework for short-term Australian mortality forecasts. These results are consistent with Giacomello \cite{Gia2024}, who applied a multivariate functional $\alpha$\nobreakdash-transformation to provincial mortality data in Italy. However, the application of the $\alpha$\nobreakdash-transformation within a CoDA framework for all-cause mortality forecasting, treating age as discrete rather than a continuum, remains unexplored. This paper addresses the gap by evaluating forecast performance across multiple countries, showing that the $\alpha$\nobreakdash-transformation can yield results comparable to or better than those obtained using the CLR transformation.

The remainder of the paper is organised as follows. Section 2 describes the methodology, outlining the analytical framework and procedures employed in this study. Section 3 presents the results, providing a detailed discussion of forecast accuracy under each transformation. Finally, Section 4 concludes the paper by summarising the main findings and highlighting potential directions for future research.


\section{Methodology}

This study involves several phases, including data preprocessing, $\alpha$\nobreakdash-parameter tuning, modelling and forecasting, and model evaluation. All analyses are conducted using R Statistical Software version 4.1.1 \cite{Rstats2024}.


\subsection{Data preprocessing}

Observed mortality data for males and females, from age 0 to the open interval 110+, are retrieved for 31 selected European countries/regions covering 1983-2018 from the Human Mortality Database (HMD) \cite{HMD2024} using the \texttt{demography} package in R \cite{Hyn2023}. These countries/regions are selected to maximise data completeness while ensuring a common timeframe. The pre-pandemic period is chosen to avoid anomalies and uncertainties caused by COVID-19, as forecast performance is sensitive to data quality and stability.

Following Bergeron-Boucher et al. \cite{Bgr2017}, observed death counts $D_{t,x}$ for each country/region are first calculated from mortality rates $m_{t,x}$ and the corresponding exposure-to-risk $E_{t,x}$. At older ages (80+), $m_{t,x}$ often exhibit considerable random variation due to unboundedly high rates, smaller denominators in $E_{t,x}$ or measurement error \cite{Protocol2021, Bas2023}. To address this, the Kannisto model \cite{Tha1998} is applied to smooth mortality rates for older ages, separately for males and females \cite{Protocol2021}. This Poisson log-likelihood approach fits a logistic curve to better capture old-age mortality patterns than alternative models \cite{Protocol2021}. As a result, zeros and missing values are eliminated at advanced ages.

For ages below 80, zeros occur in some countries/regions for specific years, which can lead to undefined results under the CLR transformation. A multiplicative replacement strategy that replaces zeros with small positive values \cite{Mar2003} is applied to impute zeros in $D_{t,x}$ \cite{Bgr2017}. This non-parametric method is coherent with simplex operations, preserves the covariance structure of non-zero components and minimally distorts the overall mortality pattern \cite{Mar2003}, while enabling log-ratio transformations. Although the $\alpha$\nobreakdash-transformation can handle zeros, replacement is applied for both transformations to ensure a fair comparison between them.

A $P$-part composition $\mathbf{r} = [r_1, r_2, \dots, r_P]$ of $D_{x}$ containing zeros is replaced by a composition $\mathbf{x} = [x_1, x_2, \dots, x_P]$ without zeros as follows \cite{Bgr2017}:
\begin{equation}
    x_i =
 \begin{cases}
     \delta, & \text{if } r_i = 0, \\
     \left( 1 - \frac{w \delta}{c} \right) r_i, & \text{if } r_i > 0,
 \end{cases}
\end{equation}
where $w$ is the number of zeros counted in $\mathbf{r}$, $c$ is the constant of the sum constraint ($\sum r_i = c$) and $\delta$ is the imputed value on part $r_i$ computed as
\begin{equation}
 \delta_{t} = \frac{\displaystyle \left(\min_{t,x} D_{t,x}\right) / 2}{\sum_{x=0}^{110} D_{t,x}}, \quad \forall \; D_{t,x} > 0.
\end{equation}
Subsequently, $\mathbf{x}$ is multiplied by $\sum_{x=0}^{110} D_{t,x}$ to generate a new set of death counts without zeros. The corresponding mortality rates for ages below 80 are then computed and combined with smoothed older-age rates to produce a complete set of age-specific mortality rates \cite{Bgr2017}.

Life tables are constructed from the preprocessed mortality rates separately for each country/region and gender using the \texttt{LifeTable} function from the \texttt{MortalityLaws} package \cite{Pas2024}. The average number of years lived by individuals dying within the age interval $[x, x + 1)$, denoted as $a_x$, is assumed to be 0.5 for all single-year ages except age 0 \cite{Bgr2017, Protocol2021}. For age 0, $a_0$ is computed for each country/region as the average of $a_0$ values derived from the range of infant mortality rates $m_0$ during the training period (1983-2010), as outlined in Table~\ref{tab:a0} \cite{And2015}. This approach follows HMD practices \cite{Protocol2021} and allows country- and gender-specific adjustments for more accurate infant mortality representation. The life table radix is assumed to be unity, ensuring that $d_{t,x}$ values fall within a standard simplex \cite{Ait1994, Oep2008, Bgr2017}. For visualisation, $d_{t,x}$ values are scaled by 100,000, a commonly used radix in demographic research \cite{Protocol2021, HLSWeighted2024, HLSAlpha2024}.

\begin{table}
\tbl{Calculation of $a_0$ based on $m_0$ \cite{And2015}.}
{\begin{tabular}{lcc} \toprule
     Gender & $m_0$ range & Formula \\ \midrule
     Male & $[0, 0.02300)$ & $\frac{1}{N} \sum_{t=1}^{N} \left( 0.14929 - 1.99545 m_{t,0} \right)$ \\
     & $[0.02300, 0.08307)$ & $\frac{1}{N} \sum_{t=1}^{N} \left( 0.02832 + 3.26021 m_{t,0} \right)$ \\
     & $[0.08307, \infty)$ & 0.29915 \\
     Female & $[0, 0.01724)$ & $\frac{1}{N} \sum_{t=1}^{N} \left( 0.14903 - 2.05527 m_{t,0} \right)$ \\
     & $[0.01724, 0.06891)$ & $\frac{1}{N} \sum_{t=1}^{N} \left( 0.04667 + 3.88089 m_{t,0} \right)$ \\
     & $[0.06891, \infty)$ & 0.31411 \\ \bottomrule
\end{tabular}}
\tabnote{\textit{Note}. $N = 28$ is the length of the training set.}
\label{tab:a0}
\end{table}


\subsection{The CLR transformation}

A $P$-part compositional vector $\mathbf{z} \in (0, \infty)^P$ can be represented as a point in the sample space, known as the simplex $\mathbb{S}^P$, through the closure operator $\mathcal{C}$, defined as
\begin{subequations} \label{subeq:simplex}
\begin{align}
    &\mathbb{S}^P = \left\{ \mathbf{x} = \left[ x_1, \dots, x_P \right] \mid x_i > 0, \; \sum_{i=1}^{P}{x_i = c} \right\}, \\
    & \mathcal{C}: (0, \infty)^P \to \mathbb{S}^P, \quad \mathcal{C}(\mathbf{z}) = \left[ \frac{z_1}{\sum_{i=1}^{P}{z_i}}, \dots, \frac{z_P}{\sum_{i=1}^{P}{z_i}} \right] = \mathbf{x},
\end{align}
\end{subequations}
where $\mathbf{x}$ is a $P$-part positive composition that carries only relative information and sums to an arbitrary constant $c$, which can be set to 1 without loss of generality \cite{Cla2022, Gia2024, Tsa2011}. Further details on CoDA operations and Aitchison geometry are provided in the Supplemental Material.

The CLR transformation \cite{Ait1986} is a one-to-one mapping between $\mathbb{S}^P$ and $(P-1)$-dimensional subspace of $\mathbb{R}^P$ under a zero-sum constraint. It preserves distances such that $d_a \langle \mathbf{x}, \mathbf{y} \rangle = d_e ( \text{clr}(\mathbf{x}), \text{clr}(\mathbf{y}))$, with $d_e$ the Euclidean distance \cite{Cla2022}. It is defined as
\begin{subequations} \label{subeq:clr}
\begin{align}
    \text{clr: } \mathbb{S}^P \to \mathbb{H} \subset \mathbb{R}^P, \quad
    & \mathbf{w} = \text{clr}(\mathbf{x}) = \left[ \ln{\frac{x_1}{g(\mathbf{x})}}, \dots, \ln{\frac{x_P}{g(\mathbf{x})}} \right], \\
    & \text{clr}^{-1} (\mathbf{w}) = \mathcal{C}(e^{\mathbf{w}}),
\end{align}
\end{subequations}
where $g(\mathbf{x}) = \left( \prod_{i=1}^{P} x_i \right)^{\frac{1}{P}}$ is the geometric mean of $\mathbf{x}$.


\subsection{The $\alpha$-transformation}

The $\alpha$\nobreakdash-transformation \cite{Tsa2011} is a one-parameter Box-Cox type power transformation that maps a $P$-part composition from $\mathbb{S}^P$ to $\mathbb{R}^{P-1}$. For any compositional vector $\mathbf{x} \in \mathbb{S}^P$, the $\alpha$-transformation is defined as
\begin{subequations} \label{subeq:alpha}
\begin{align}
    A_\alpha: \mathbb{S}^P \to \mathbb{R}^{P-1}, \quad
    & \mathbf{z}_\alpha = A_\alpha (\mathbf{x}) = \frac{1}{\alpha} \mathbf{H}_P \left( P \mathbf{u}_\alpha (\mathbf{x}) - \mathds{1}_P \right), \\
    & A^{-1}_{\alpha} (\mathbf{z}_\alpha) = \mathcal{C} \left( (\alpha \mathbf{H}_{P}^{'} \mathbf{z}_\alpha + \mathds{1}_P)^{1/\alpha} \right),
\end{align}
\end{subequations}
where $\mathbf{z}_\alpha \in \mathbb{R}^{P-1}$, $\mathbf{H}_P$ is the $(P-1) \times P$ Helmert sub-matrix, $\mathbf{u}_\alpha (\mathbf{x}) = \mathcal{C}(\mathbf{x}^{\alpha})$ is the $\alpha$-transformed composition and $\mathds{1}_P$ is the $P$-dimensional vector of ones. By removing the first row of the orthonormal Helmert matrix, $\mathbf{H}_P$ provides an orthogonal basis on $\mathbb{R}^{P}$ and reduces the dimensionality of the transformed composition to $P-1$ \cite{Lan1965}.

The parameter $\alpha \in [0,1]$ can be tuned using criteria appropriate for the application, such as pseudo-$R^2$, profile log-likelihood or the Kullback–Leibler divergence \cite{Tsa2011, Tsa2015}. When $\alpha \to 1$, the transformation simplifies to a linear transformation, whereas $\alpha = 0$ is equivalent to the isometric log-ratio (ILR) transformation \cite{Tsa2011}:
\begin{equation}
    \text{ilr: } \mathbb{S}^P \to \mathbb{R}^{P-1}, \quad \text{ilr}(\mathbf{x}) = \mathbf{H}_P \; \text{clr} (\mathbf{x}). 
\end{equation}


\subsection{Modelling and forecasting}

Following established procedures in \cite{Bgr2017, Oep2008}, the CoDA model is formulated as
\begin{equation}
    A_\alpha(d_{t,x} \ominus \alpha_x) = \kappa_t \beta_x + \varepsilon_{t,x}, 
\end{equation}
where $\alpha_x$ is the age-specific geometric mean, $\varepsilon_{t,x}$ denotes the error and $\ominus$ is the negative perturbation. For each country/region and gender, a matrix $\mathbf{D}$  of $d_{t,x}$ is constructed, with $T$ rows representing the number of years and $X+1$ columns representing the ages $x$. In this study, the data are split into training and test sets using the commonly chosen 80:20 ratio \cite{Hyn2021}, resulting in a training period from 1983 to 2010 and a test period from 2011 to 2018. Accordingly, each training matrix $\mathbf{D}$ has 28 rows (1983–2010) and 111 columns (ages 0 to 110+), with each row summing to the life table radix.

The matrix $\mathbf{D}$ is first centred by $\alpha_x$ to obtain matrix $\mathbf{F}$. Transformations are then applied to $\mathbf{F}$ to map the compositional data to real space, yielding the matrix $\mathbf{H}$. Subsequently, SVD is applied to $\mathbf{H}$ to estimate $\kappa_t$ and $\beta_x$ via a rank-$k$ approximation, resulting in a low-rank matrix $\mathbf{H}^*$ to be forecasted. While a rank-1 approximation is common, higher ranks are used when the first component does not explain sufficient variance \cite{Bgr2017}. Based on the proportion of explained variance, $k = 7$ for females and $k = 4$ for males are deemed appropriate, as they each account for over 80\% of the total variance on average. This leads to $k$ series of estimated $\kappa_t$ for each dataset.

Each $\kappa_t$ series is forecasted using an autoregressive integrated moving average (ARIMA) model over an 8-year horizon. The approximately linear trend of $\kappa_t$ makes ARIMA model appropriate. Although a random walk with drift often provides good fit \cite{Lee1992}, prior work \cite{Bgr2017} found ARIMA (0,1,1) with drift performs well for most Western European countries. Therefore, this study considers two forecasting models, namely (i) the default ARIMA (0,1,1) with drift and (ii) the automatic ARIMA model which selects the optimal order via a stepwise algorithm \cite{Hyn2008}. 

The resulting matrix is subsequently inverse-transformed to obtain $\mathbf{F}^*$ in the simplex. The $\alpha_x$ values are then compositionally added back and a jump-off adjustment is applied, yielding the matrix $\mathbf{D}^*$ which contains the forecasted life table death counts $\hat{d}_{t,x}$. Finally, a bias-corrected back-transformation to the original scale, as described in Algorithm 1 of the Supplemental Material, is applied to ensure demographically meaningful forecasts of $d_{t,x}$. A complete life table can then be constructed from $\hat{d}_{t,x}$, thereby providing the full age-specific mortality profile of the population.


\subsection{Model evaluation}

Forecast accuracy of each model is evaluated on the test set using root mean squared error (RMSE) and mean absolute error (MAE). The ARIMA specification with the lowest out-of-sample error is selected as the best model. A comparative analysis is subsequently conducted to assess the impact of the $\alpha$\nobreakdash-transformation on forecast performance relative to the benchmark CLR transformation.


\section{Application to real data}

This section presents and discusses the forecast performance of the CoDA models under both transformations.


\subsection{$\alpha$-parameter tuning}

The values of $\alpha$ are chosen on a data-driven basis via cross-validation \cite{Tsa2015}. In order to determine the optimal $\alpha$ values for transforming $d_{t,x}$ in each country/region, an expanding window approach \cite{Hyn2021} is adopted. As aforementioned, the training set spans 1983-2010, with the remaining eight years (2011-2018) reserved as the test set. 

Within the training set, an additional split into sub-training and validation sets is performed. Starting with an initial sub-training period of 15 years, the window expands by one year at a time while maintaining a fixed four-year validation period, resulting in ten iterations. For instance, in the first iteration, the sub-training set spans 1983-1997 with validation covering 1998-2001. In the last iteration, sub-training spans 1983-2006 with validation covering 2007-2010. This progressive expansion improves generalisability and mitigates overfitting to limited data.

The $\alpha$\nobreakdash-parameter offers flexibility to adapt to different mortality patterns, but excessive flexibility may cause overfitting if $\alpha$ is overly sensitive to small variations. To address this, the \texttt{optim} function in R \cite{Rstats2024} is used to select optimal $\alpha$ values within the range of $[0,1]$ that minimise the average RMSE in the validation set. A penalisation mechanism is also applied to exclude $\alpha$ values that produce implausible $\hat{d}_{t,x}$.

Using female data as an example, Table~\ref{tab:female-alpha} reports the optimal $\alpha$ values and their corresponding average validation RMSE. Interestingly, only five countries/regions have an optimal $\alpha$ value of 0, indicating that the intermediate $\alpha$ values are generally more suitable for female mortality forecasting than LRA.

\begin{table}
\tbl{Optimal $\alpha$ values for female mortality data.}
{\begin{minipage}[t]{0.46\textwidth}
    \centering
    \begin{tabular}{lcc} \toprule
    Country/Region & Optimal $\alpha$ & Avg. RMSE (\%) \\ \midrule
    Austria & 0.2465 & 0.0495 \\
    Belgium & 0.9901 & 0.0555 \\
    Bulgaria & 0.1000 & 0.0728 \\
    Belarus & 0.1000 & 0.0999 \\
    Switzerland * & 0.0000 & 0.0553 \\
    Czechia & 0.3338 & 0.0593 \\
    East Germany & 0.3235 & 0.0681 \\
    West Germany & 0.8124 & 0.0450 \\
    Denmark & 0.1499 & 0.0992 \\
    Spain * & 0.0000 & 0.0441 \\
    Estonia & 0.1808 & 0.1320 \\
    Finland & 0.0852 & 0.1102 \\
    France & 0.1994 & 0.0588 \\
    England \& Wales * & 0.0000 & 0.0501 \\
    Northern Ireland & 0.1000 & 0.1108 \\
    Scotland & 0.1000 & 0.0653 \\ \bottomrule
\end{tabular}
\end{minipage}
\hspace{0.8cm}
\begin{minipage}[t]{0.46\textwidth}
    \centering
    \begin{tabular}{lcc} \toprule
    Country/Region & Optimal $\alpha$ & Avg. RMSE (\%) \\ \midrule
    Greece & 0.0044 & 0.0617 \\
    Hungary & 0.2231 & 0.0691 \\
    Ireland & 0.1000 & 0.1182 \\
    Iceland & 0.2699 & 0.2668 \\
    Italy & 0.5148 & 0.0404 \\
    Lithuania & 0.1000 & 0.0903 \\
    Luxembourg & 0.2637 & 0.1993 \\
    Latvia & 0.1000 & 0.0893 \\
    Netherlands & 0.1000 & 0.0716 \\
    Norway & 0.1000 & 0.0731 \\
    Poland * & 0.0000 & 0.0648 \\
    Portugal * & 0.0000 & 0.0706 \\
    Slovakia & 0.1000 & 0.0733 \\
    Slovenia & 0.1000 & 0.0926 \\
    Sweden & 0.0999 & 0.0657 \\ 
    ~ & ~ & ~ \\ \bottomrule
\end{tabular}
\end{minipage}}
\tabnote{\textit{Note}. Asterisks (*) indicate cases where optimal $\alpha = 0$.}
\label{tab:female-alpha}
\end{table}


\subsection{Forecasts of mortality: A case study on Italian female mortality}

The results show that the $\alpha$\nobreakdash-transformation performs comparably to or better than the CLR transformation in 20 countries/regions for females and 22 for males. Using Italian female data as an example, the forecasted $\hat{d}_{t,x}$ for the last holdout year (2018), shown in Figure~\ref{fig:female-2018-ITA}, demonstrates a negatively skewed, bimodal age-at-death distribution, with an initial peak in infancy shifting towards older ages \cite{HLSWeighted2024, Rob2012}. Infant mortality is primarily driven by genetic or infectious factors, whereas ageing dominates at older ages \cite{Rob2012}. According to Abouzahr et al. \cite{Abo2012}, mortality rates are typically high in infancy, reach a minimum between ages 5 and 14, and then rise exponentially after age 35. Additional bumps are occasionally observed during female reproductive ages, reflecting maternal mortality.

On the other hand, Figure~\ref{fig:female-age-ITA} shows the forecasted $\hat{d}_{t,x}$ of Italian females for selected 20-year-gapped ages over time. Death counts generally decline for infants and younger age groups up to age 80, whereas ages 100 and above show an increasing trend, reflecting population ageing. Both transformations broadly replicate historical patterns, but the $\alpha$\nobreakdash-transformation yields smoother and more stable forecasts, particularly for age groups with low or volatile death counts, such as ages 20 and 110+. In contrast, the CLR transformation appears more sensitive to variability in these ages, likely due to the amplifying effect of log-ratios.

In summary, the $\alpha$\nobreakdash-transformation produces lower forecast errors than the CLR transformation across the test period and substantially outperforms it at ages above 75, as illustrated in Figures~\ref{fig:female-error-year-ITA} and \ref{fig:female-error-age-ITA}. Additional results for both females and males are provided in the Supplemental Material.

\setcounter{figure}{0}
\begin{sidewaysfigure}
    \centering
    \subfloat[\label{fig:female-2018-ITA}]{%
        \makebox[0.48\linewidth]{
        \includegraphics[width=0.48\linewidth, height=0.3\textheight]{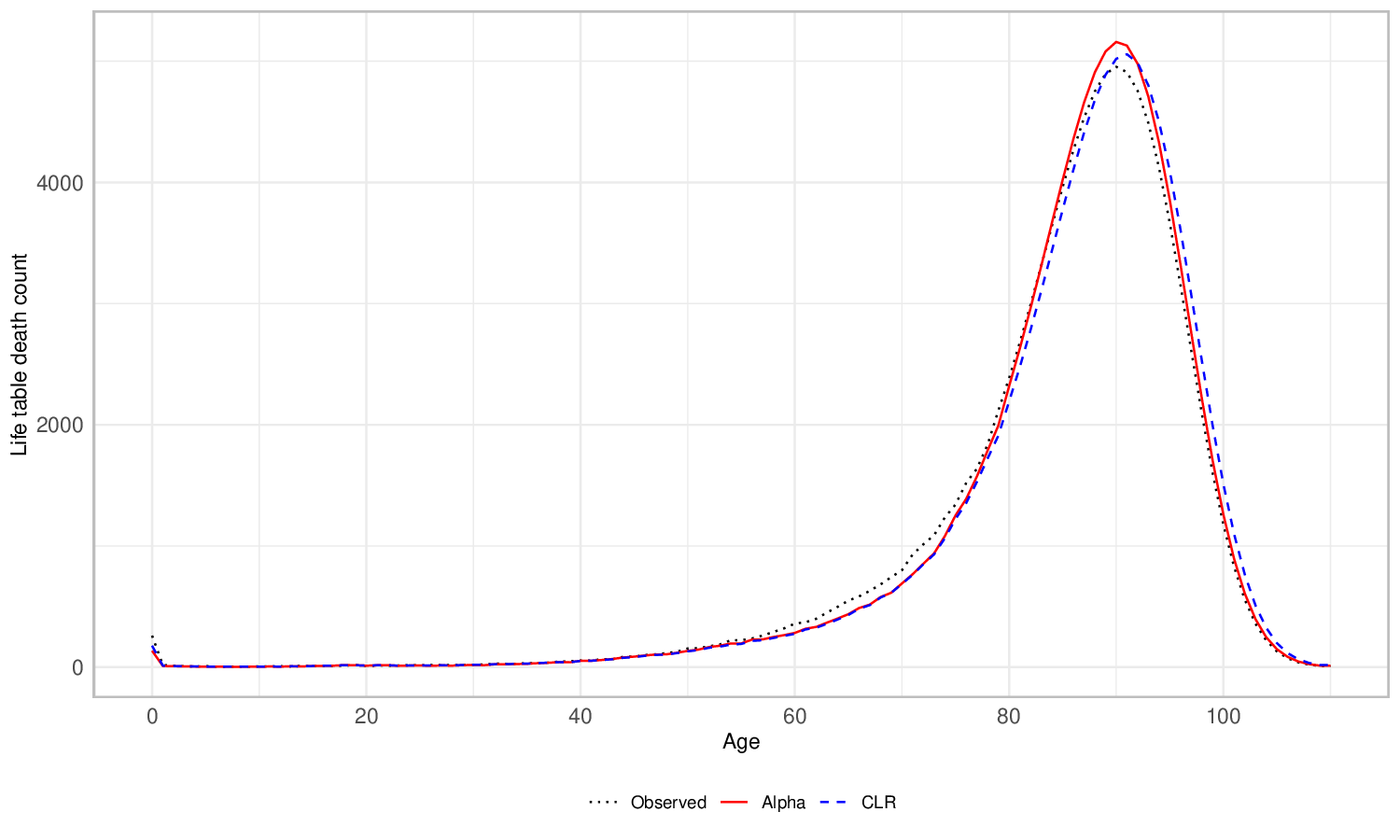}}%
    } \hfill
    \subfloat[\label{fig:female-age-ITA}]{%
        \makebox[0.48\linewidth]{
        \includegraphics[width=0.48\linewidth, height=0.3\textheight]{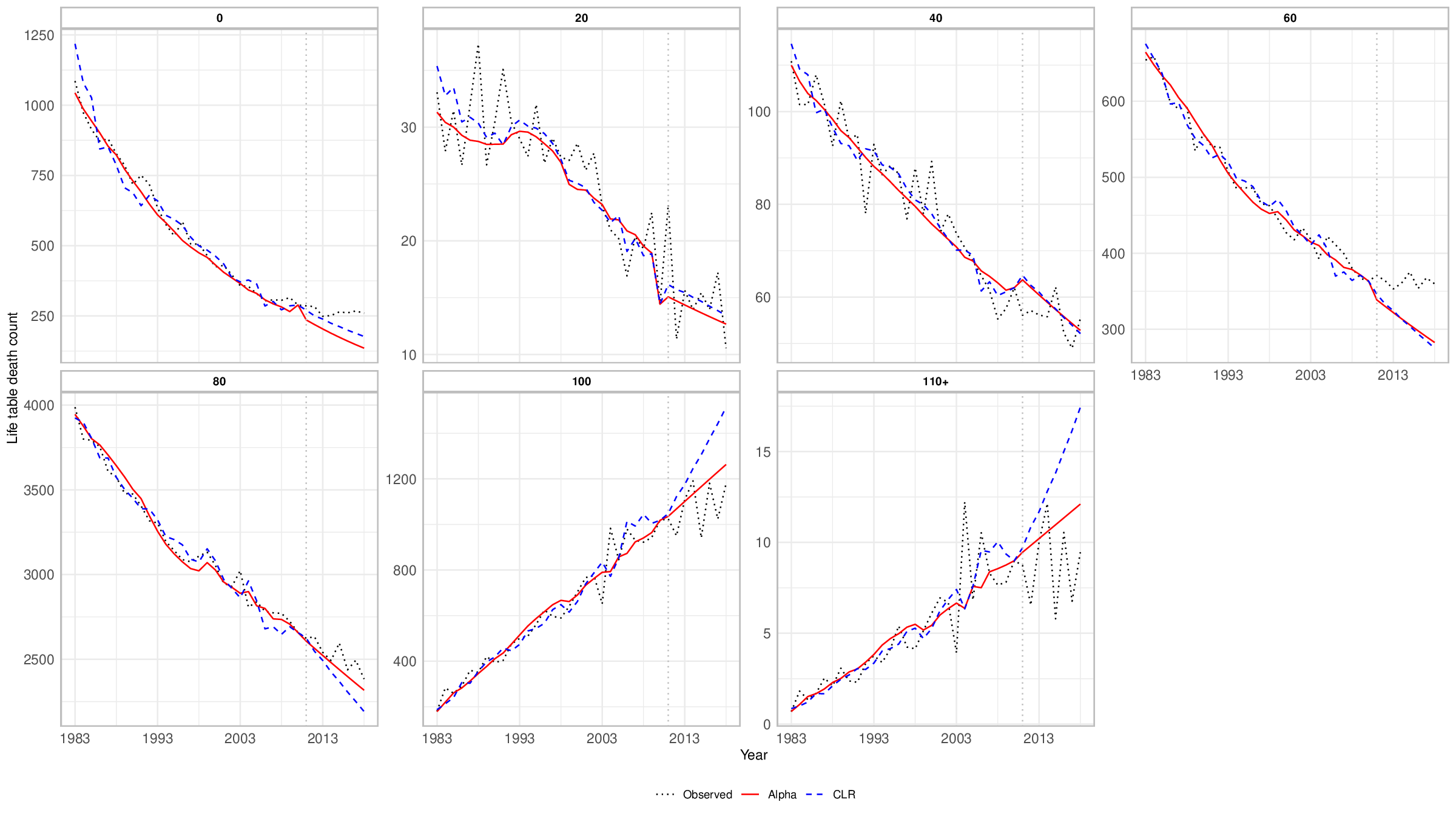}}%
    } \\
    \subfloat[\label{fig:female-error-year-ITA}]{%
        \makebox[0.48\linewidth]{
        \includegraphics[width=0.48\linewidth, height=0.3\textheight]{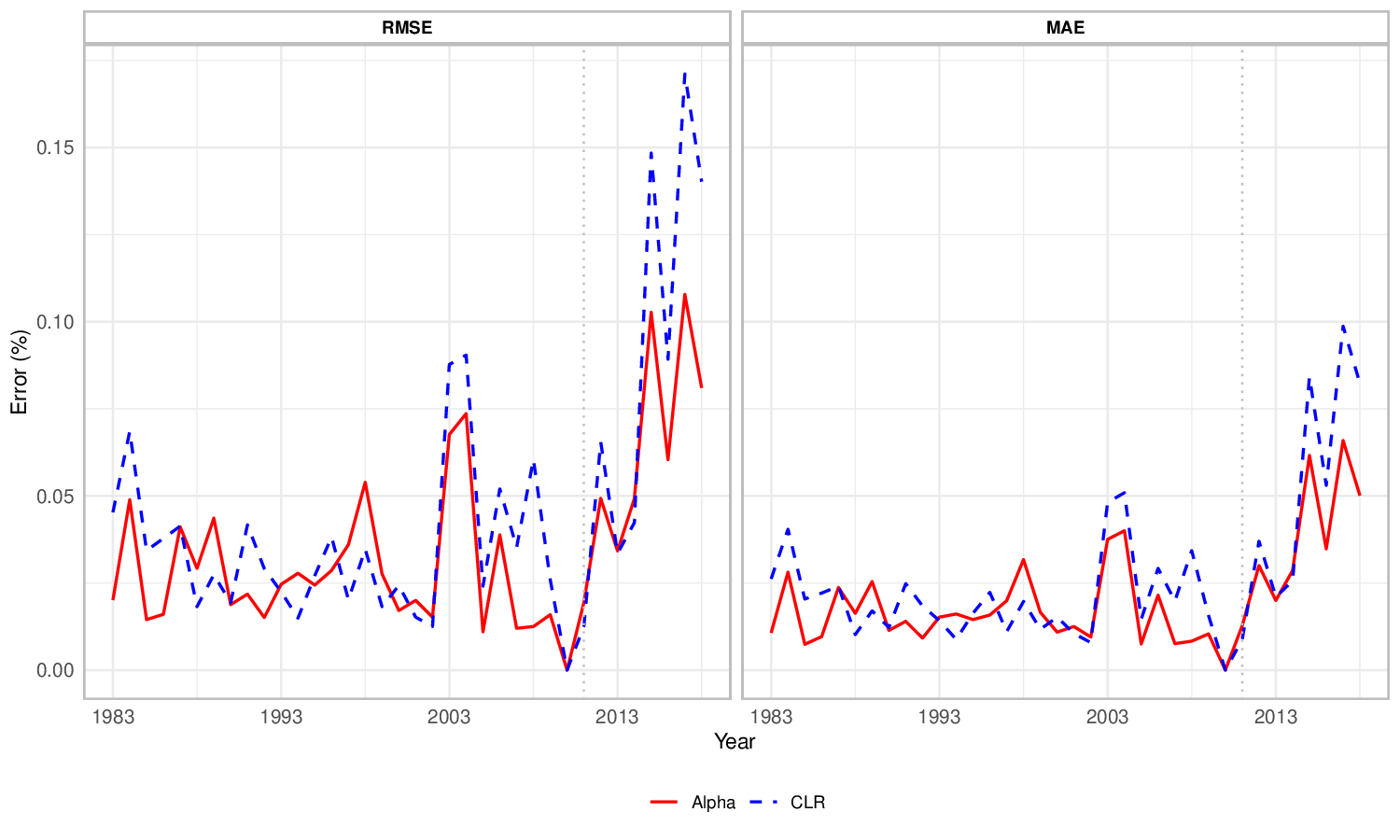}}%
    } \hfill
    \subfloat[\label{fig:female-error-age-ITA}]{%
        \makebox[0.48\linewidth]{
        \includegraphics[width=0.48\linewidth, height=0.3\textheight]{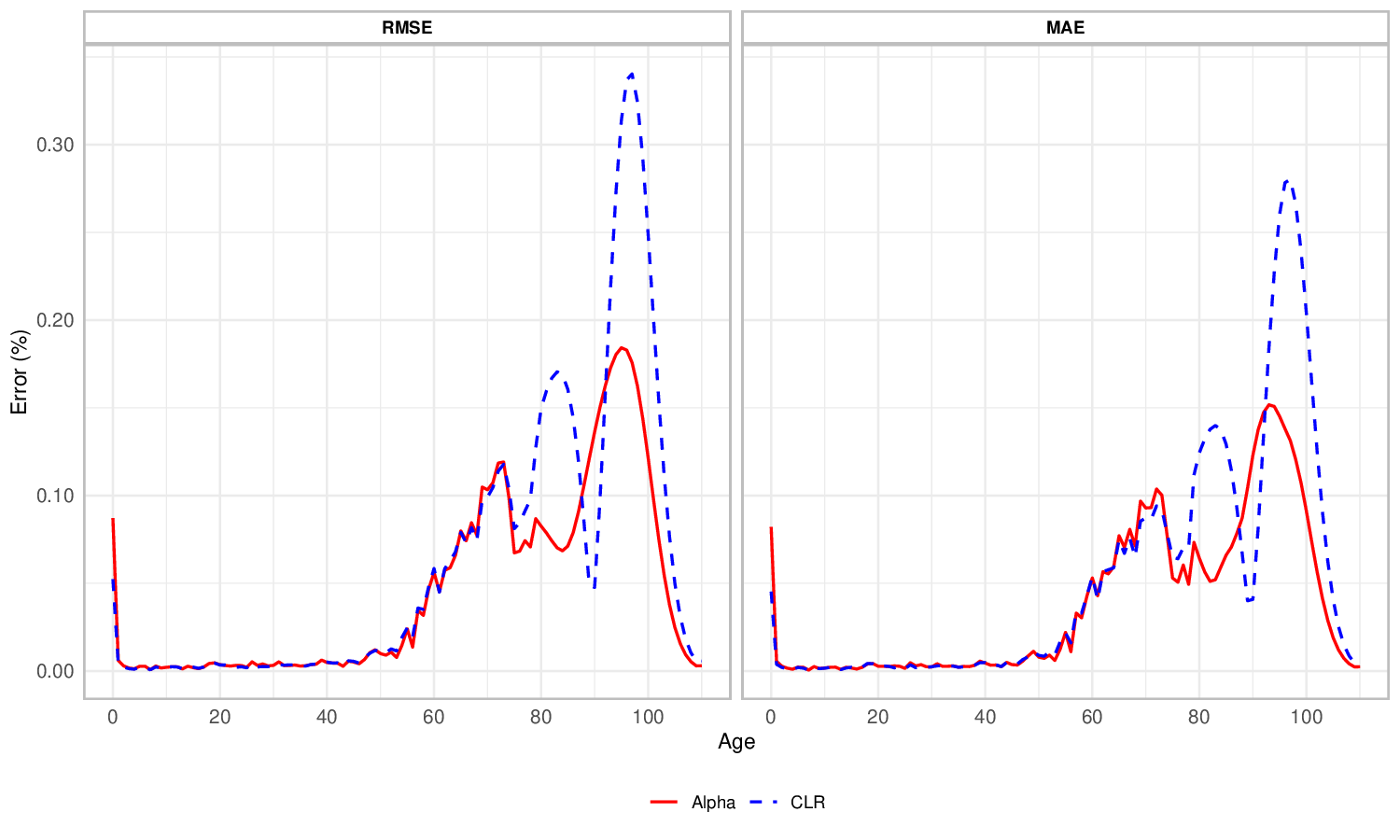}}%
    }
    \caption{Forecasted age-specific life table death counts (a) in 2018 and (b) for selected ages, with forecast errors (c) over years and (d) by ages for Italian females.}
    \label{fig:female-ITA}
\end{sidewaysfigure}


\subsection{Comparison of mean forecast accuracy}

\subsubsection{Overall mean errors}

Table~\ref{tab:female-error-overall} summarises the overall mean forecast errors for female mortality in both the training and test sets. The two transformations perform comparably across both sets, with the $\alpha$\nobreakdash-transformation yielding slightly lower errors. Similar findings have been reported by Giacomello \cite{Gia2024} and Shang and Haberman \cite{HLSAlpha2024}. The improved performance is primarily due to the flexibility of the $\alpha$\nobreakdash-parameter, which enables the transformation to better capture the underlying data structure \cite{Tsa2011}, particularly temporal changes in age-specific life table death counts \cite{HLSAlpha2024}.

\begin{table}
\tbl{Overall mean forecast errors for female mortality.}
{\begin{tabular*}{\textwidth}{@{\extracolsep{\fill}}lcccc} \toprule
     & \multicolumn{2}{c}{RMSE (\%)} & \multicolumn{2}{c}{MAE (\%)} \\ \cmidrule(l){2-3} \cmidrule(l){4-5}
     Phase & $\alpha$ & CLR & $\alpha$ & CLR \\ \midrule
     Train & \textbf{0.0614} & 0.0616 & \textbf{0.0346} & 0.0347 \\
     Test & \textbf{0.0880} & 0.0909 & \textbf{0.0522} & 0.0534 \\ \bottomrule
\end{tabular*}}
\label{tab:female-error-overall}
\end{table}


\subsubsection{Mean errors by country/region}

\begin{figure}
    \centering
    \includegraphics[width=1\linewidth]{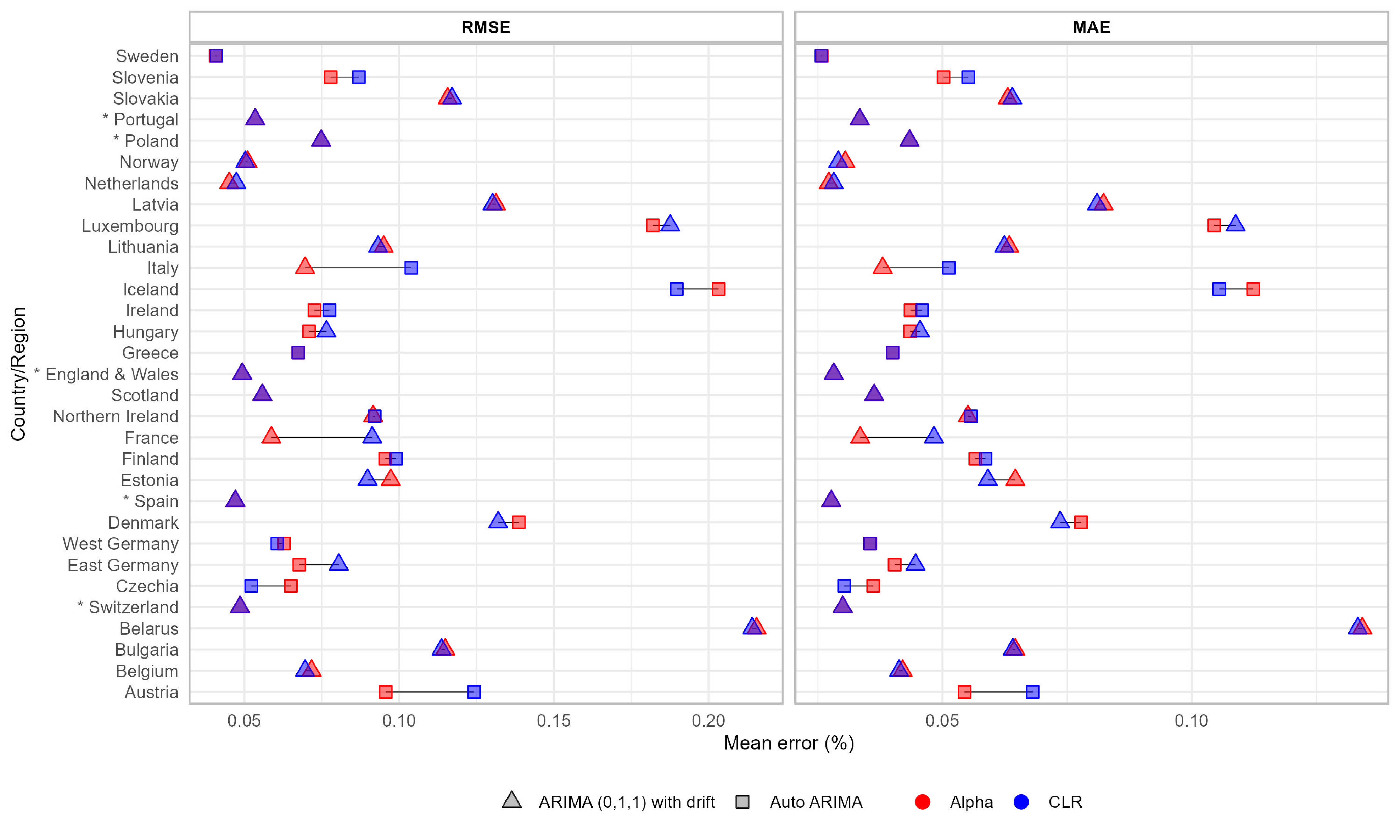}
    \caption{Mean forecast errors of the best ARIMA models for female mortality in each country/region. Asterisks (*) indicate cases where optimal $\alpha = 0$.}
    \label{fig:female-error-test}
\end{figure}

Figure~\ref{fig:female-error-test} presents the mean forecast errors of the best ARIMA models for female mortality in each country/region. Notably, the ARIMA (0,1,1) with drift appears as the best-performing model in more than half of the cases, consistent with findings reported in previous study \cite{Bgr2017}. In general, the $\alpha$\nobreakdash-transformation yields comparable or lower forecast errors than the CLR transformation in 20 countries/regions for females. Many cases where the $\alpha$\nobreakdash-transformation underperforms are generally populations where the first $k = 7$ singular values explain less than 80\% of the total variation and the population is relatively small, leading to higher stochastic variability and reduced data stability. It is also worth noting that when the optimal $\alpha = 0$, as observed in Switzerland, Spain, England \& Wales, Poland and Portugal, the transformation reduces to the ILR transformation and provides accuracy similar to the CLR transformation. This aligns with Shang and Haberman \cite{HLSAlpha2024}, who found that ILR and CLR transformations yield broadly comparable point forecast accuracy for Australian gender-specific $d_{t,x}$ within the functional CoDA framework.

\subsubsection{Mean errors over years}

\begin{figure}
    \centering
    \includegraphics[width=0.9\linewidth]{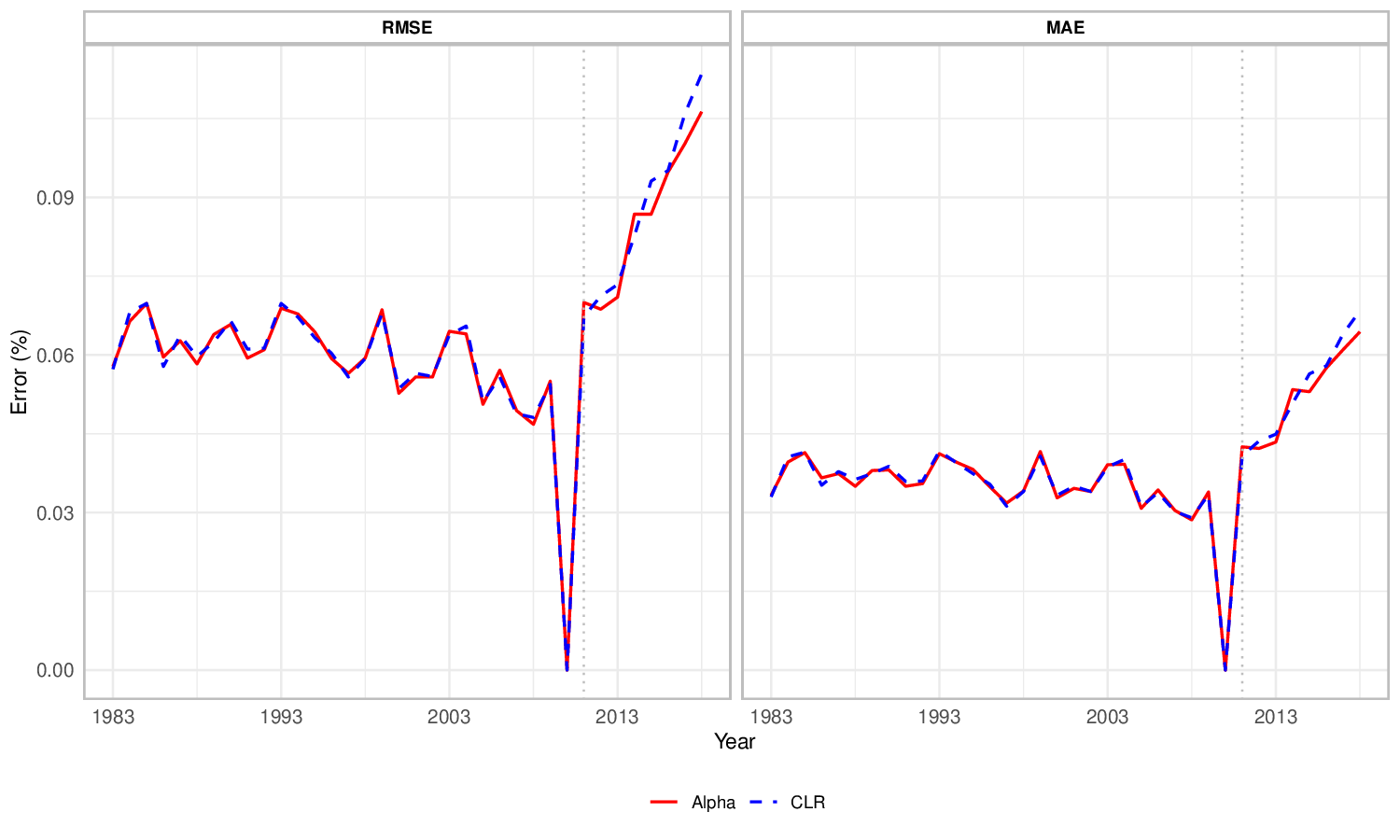}
    \caption{Mean forecast errors for female mortality over years across countries/regions.}
    \label{fig:female-error-year}
\end{figure}

Figure~\ref{fig:female-error-year} visualises the trend of mean forecast errors for female mortality over years in the training and test sets. The $\alpha$\nobreakdash-transformation performs comparably to the CLR transformation with slightly lower errors, particularly in the later years of the forecast horizon. In addition, the extremely small errors in the last fitting year (2010) are due to the jump-off adjustment, which uses the last observed $d_{t,x}$ as the jump-off point to ensure continuity between observed and forecasted life table death counts \cite{Lee2001, Sto2018}.

When the classical LC model was introduced, Lee and Carter \cite{Lee1992} noted that using fitted values as jump-off rates may not perfectly align with the observed mortality in the jump-off year. This mismatch can create a discontinuity between observed and forecasted log mortality rates, particularly affecting ages with very low mortality. Later, Lee and Miller \cite{Lee2001} emphasised that this jump-off bias can substantially reduce short-term forecast accuracy, especially when low-rank approximations introduce additional error \cite{Bell1997}. Both Bell \cite{Bell1997} and Lee and Miller \cite{Lee2001} showed that adjusting forecasts to match observed jump-off rates effectively eliminates this bias and improves forecast accuracy, a finding further confirmed by Stoeldraijer et al. \cite{Sto2018}. Consistent with this approach, the errors between $d_{t,x}$ and $\hat{d}_{t,x}$ reported here are negligible, as the fitted values are explicitly adjusted to align with the observed $d_{t,x}$ in the jump-off year \cite{Lee2001}.


\subsubsection{Mean errors by age}

\begin{figure}
    \centering
    \includegraphics[width=0.9\linewidth]{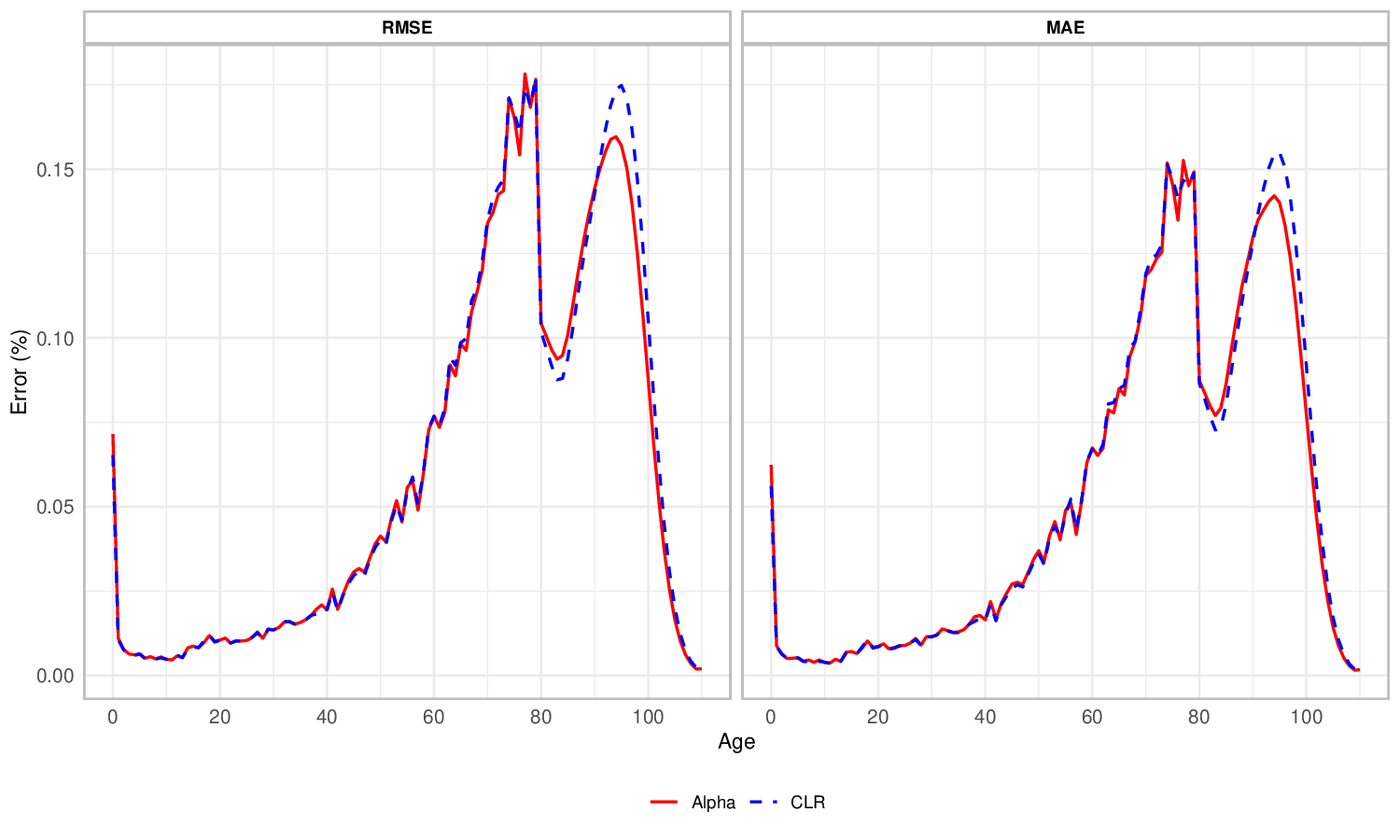}
    \caption{Mean forecast errors for female mortality by age across countries/regions.}
    \label{fig:female-error-age}
\end{figure}

Figure~\ref{fig:female-error-age} illustrates the mean forecast errors for female mortality by age across countries/regions. Forecast accuracy is broadly comparable between the $\alpha$\nobreakdash- and CLR-transformed data. While the $\alpha$\nobreakdash-transformation exhibits slightly higher errors around ages 80 to 90, it shows a clear advantage over the CLR transformation at ages above 90. These findings align with previous findings, suggesting that the $\alpha$\nobreakdash-transformation represents a viable alternative to the CLR transformation in compositional mortality forecasting \cite{Gia2024, HLSAlpha2024}.

In addition to the results reported here, the Supplemental Material provides a comparison of forecast performance between the CoDA and LC models, together with an evaluation of the $\alpha$\nobreakdash-transformed CoDA model using zero-containing data. To explore model behaviour under extreme mortality conditions, the study period is extended to 2022 to include the COVID-19 pandemic years (2020-2022) across 23 countries/regions. Forecasts over the COVID-19 period are generated fully out of sample, with the pandemic years included exclusively in the test set and thus no shock-related information available during model training. Under these extreme mortality shocks, the $\alpha$\nobreakdash-transformed CoDA approach exhibits less systematic overforecasting relative to competing models.


\section{Conclusion}

This study investigates the use of the $\alpha$\nobreakdash-transformation as an alternative to the commonly used CLR transformation within the CoDA framework for discrete-age mortality forecasting. Unlike the classical LC model, life table death counts are employed as the mortality measure. Due to their compositional nature, these data are transformed into real space prior to forecasting. Using age-specific life table death counts across selected European countries/regions for males and females, a comparative analysis of the forecast performance of the two transformations is conducted.

The results indicate that models fitted to the $\alpha$\nobreakdash-transformed data perform comparably to those using the CLR transformation across most countries/regions, with improved forecast accuracy observed in certain cases. This is consistent with previous studies within the functional CoDA framework \cite{Gia2024, HLSAlpha2024}. The advantage of the $\alpha$\nobreakdash-transformation is particularly pronounced at older ages, where it yields noticeably improved accuracy for ages with low and volatile death counts. This improvement is primarily attributable to the flexibility of the $\alpha$\nobreakdash-parameter which bridges EDA and LRA \cite{Tsa2016}, allowing for better adaptation to the data. In this study, optimal $\alpha$ values for most datasets fall within the intermediate range, determined by minimising the average validation RMSE during parameter tuning.

A key limitation of this study arises from the dataset itself. Forecast accuracy is highly data-dependent, with factors such as country selection, the length of the fitting period and the granularity of age groupings potentially affecting model performance. High-quality datasets of sufficient size are therefore essential to ensure robust and reproducible results. Future research could investigate alternative power transformations that accommodate zero values, such as the chiPower transformation \cite{Gre2024}, to further improve forecast accuracy. Extensions may also examine the performance of different transformations within a functional coherent CoDA framework to capture common trends across countries. From an actuarial perspective, including applications in the pension and insurance sectors, forecasts of life table death counts can be used to estimate annuity prices, as demonstrated in the literature \cite{HLSWeighted2024}.


\section*{Acknowledgements}

The authors acknowlegde the use of ChatGPT (OpenAI, GPT-5; accessed November 2025) for language polishing, summarising explanatory text and improving the readability of pre-existing code. All methods, analyses and code logic are the authors’ own and all AI-assisted material was checked and edited by the authors.


\section*{Disclosure statement}

No potential conflict of interest was reported by the author(s).


\section*{Funding}

This work was supported by the Universiti Malaya Research Excellence Grant (UMREG 2.0) 2024 [UMREG044-2024].


\section*{Data availability statement}

All data used in this manuscript are available from the Human Mortality Database (www.mortality.org).


\bibliographystyle{tfs}
\bibliography{interacttfssample}

\end{document}